\documentclass[]{spie}  
\usepackage{amsmath,amsfonts,amssymb}
\usepackage[colorlinks=true, allcolors=blue]{hyperref}

\usepackage[]{graphicx}
\usepackage{multirow}
\usepackage{array}
\usepackage{amsmath}
\usepackage{graphicx}
\usepackage{dcolumn}
\usepackage{bm}
\usepackage{verbatim}
\DeclareMathAlphabet \mathbfcal{OMS}{cmsy}{b}{n}
\title{Laser pulse waveform control of Dirac fermions in graphene} 

\author{S. Azar Oliaei Motlagh, Vadym Apalkov, and Mark I. Stockman
\skiplinehalf
Center for Nano-Optics (CeNO) and
Department of Physics and Astronomy, Georgia State
University, Atlanta, Georgia 30303, USA
}

\authorinfo{Further author information: (Send correspondence to S.A.O.M.)\\S.A.O.M.: E-mail: soliaeimotlagh1@gsu.edu \\  V.A.: E-mail: vapalkov@gsu.edu\\  M.I.S: E-mail: mstockman@gsu.edu}

\begin{document} 
\maketitle

\begin{abstract}
We theoretically study the Dirac fermion dynamics in a graphene monolayer in the presence of an applied ultrafast laser pulse. The  pulse has the duration of a few femtoseconds and the amplitude of ∼ 0.1 - 0.5 $\mathrm{V/\AA}$. The waveform of the pulse is described by Hermit Gaussian polynomials with varying carrier-envelope phase. We show that the ultrafast dynamics of Dirac fermions strongly depends on the carrier-envelope phase and the frequency of the applied pulse. The ultrafast pulse generates an  electric current which results in a finite transferred charge. The ultrafast field-driven current and the corresponding net transferred charge depend on the waveform of the applied pulse. Our results pave the way for the development of ultrafast information processing in the terahertz domain.
\end{abstract}



\section{Introduction}
	
The interaction of an ultrafast laser pulse with solids creates unique platforms to control the electron motion with high temporal resolution. The availability of the laser pulses with few femtoseconds duration and very high intensity comparable with the internal forces acting on the electron in solids provides great tools to scientists to study the nonlinear and ultrafast electron dynamics inside the materials \cite{Gudde_et_al_Progress_Surface_Science_2005_tr_image,Schiffrin_at_al_Nature_2012_Current_in_Dielectric, Apalkov_Stockman_PRB_2012_Strong_Field_Reflection, Stockman_et_al_PhysRevB.95_2017_Crystalline_TI,Stockman_et_al_PhysRevB.98_2018_3D_TI,Stockman_et_al_PhysRevB.99_2019_Weyl, Ghimire_et_al_Nature_Communications_2017_HHG, Reis_et_al_Nat_Phys_2017_HHG_from_2D_Crystals, Simon_et_al_PRB_2000_Strong_Field_Fs_Ionization_of_Dielectrics,  Mashiko_et_al_Nature_Communications_2018_ultrafast_pulse_solid, Shin_et_al_IOP_Publishing_2018_ultrafast_pulse_solid,  Sun_et_al_Chinese_Physics_B_2017_Ultrafast_pulses_TMDC, Zhang_et_al_OSA_2018_ultrafast_pulse_TMDC,Stockman_et_al_12.2506689_2019_Weyl_semimetals,Oliaei_Motlagh_et_al_2017_TI}. In recent years, there have been done several experimental and theoretical works on studying ultrafast electron dynamics in two dimensional (2D) materials using ultrafast pulses \cite{Gudde_et_al_Nature_10.1038_2018_Dirac_current,Stockman_et_al_PhysRevB.98_2018_Rapid_Communication_Topological_Resonances,Stockman_PhysRevB.92_2015_Ultrafast_Control_Symmetry, Higuchi_et_al_Nature_2017,Stockman_et_al_PhysRevB.97.035407_2018_Phosphorene,Stockman_et_alCLEO_AT.2019_MoS2_monolayer}.

Graphene, a very well known 2D material, has been a subject of intensive research in nonlinear and ultrafast optics due to its special optical and electrical properties \cite{ Lefebvre_et_al_JOSAB.35.000958_2018_graphene_CEP,Higuchi_Hommelhoff_et_al_Nature_2017_Currents_in_Graphene, Hommelhoff_et_al_PhysRevLett.121_2018_Coherent,Stockman_et_al_PhysRevB.96_2017_Berry_Phase,Leitenstorfer_et_al_PhysRevB.92_2015_Ultrafast_Pseudospin_Dynamics_in_Graphene,Hommelhoff_et_al_1903.07558_2019_laser_pulses_graphene, sun_et_al_nnano.2011.243_2012_Ultrafast_pulses_graphene,Stockman_et_al_PhysRevB.96_2017_Berry_Phase,Gruber_et_al_ncomms13948_2016_Ultrafast_pulses_graphene,Ahmadi_et_al_International_Journal_Modern_PhysB_graphene_massive}.
Graphene, a layer of carbon atoms, has a honeycomb lattice structure made of two nonequivalent triangular sublattices A and B. Having two inequivalent Dirac points at the edges of its Brillouin zone (BZ) with linear low energy dispersion relation make graphene an excellent platform for studying the dynamics of massless Dirac fermions \cite{Novoselov_et_al_Nature_Materials_2007_Rise_of_Graphene,Novoselov_Geim_et_al_nature04233_2D_Electrons_in_Graphene}. Graphene presents interesting topological characteristics like nonzero Berry connection in the vicinity of its Dirac points which leads to the nonzero Berry curvature. The Berry curvature of graphene is localized as a delta function at the Dirac points \cite{Xiao_Niu_RevModPhys.82_2010_Berry_Phase_in_Electronic_Properties}. 

In this article, we apply an ultrashort pulse with a few femtoseconds duration and several tenths of $\mathrm{V/\AA} $ amplitude normal to the plane of pristine graphene. Due to zero energy gap and strong interband coupling at the Dirac points, $K$ and $K^\prime$, very strong mixing of the valence band (VB) and the conduction band (CB) states happens near the Dirac points in the presence of the external pulse. We study the effect of ultrafast laser pulse waveform on the dynamics of an electron in the first BZ of graphene numerically. Our applied ultrafast pulse generates an ultrafast charge current and as a result a net transferred charge. We show the possibility of controlling the generated current, and the transferred charge with the waveform of the applied linearly polarized pulse. The waveform of the pulse is described  by different Hermit Gaussian expressions and is also determined by the carrier-envelope phase.

\section{MODEL AND MAIN EQUATIONS}
\label{Model_and_Equations}

We apply an ultrafast pulse with less than 5 fs duration normally to the plane of pristine graphene and we study the effects of pulse waveform on ultrafast electron dynamics. The field driven electron dynamics is coherent since the electron scatering time is longer than 10 fs \cite{Hwang_Das_Sarma_PRB_2008_Graphene_Relaxation_Time, Breusing_et_al_Ultrafast-nonequilibrium-carrier-dynamics_PRB_2011, theory_absorption_ultrafast_kinetics_graphene_PRB_2011, Ultrafast_collinear_scattering_graphene_nat_comm_2013, Gierz_Snapshots-non-equilibrium-Dirac_Nat-Material_2013, Nonequilibrium_dynamics_photoexcited_electrons_graphene_PRB_2013}  which is much longer than the duration of the pulse. To find the coherent electron dynamics in graphene we solve time-dependent Schr\"odinger equation (TDSE) 
\begin{equation}
i\hbar \frac{{d\Psi }}{{dt}} = { H(t)} \Psi,  
\label{Sch}
\end{equation}
with the Hamiltonian
\begin{equation}
{ H}(t) = { H}_0 - e{\bf{F}}(t){\bf{r}},
\label{Ht}
\end{equation}  
where $e$ is the electron charge, and $H_0$ is the  tight binding Hamiltonian of the pristine graphene \cite{Electronic_properties_graphene_RMP_2009},
\begin{eqnarray}
H_0=\left( {\begin{array}{cc}
   0 & \gamma f(\mathbf k) \\
   \gamma f^\ast(\mathbf k) & 0 \\
  \end{array} } \right) ,
\label{H0}
\end{eqnarray}
where $\gamma= -3.03$ eV is hopping integral,
\begin{equation}    
f(\mathbf k)=\exp\Big(i\frac{ak_y}{\sqrt{3}}\Big )+2\exp\Big(-i\frac{ak_y}{2\sqrt{3}}\Big )\cos{\Big(\frac{ak_x}{2}\Big )},
\end{equation}
and $a=2.46~\mathrm{\AA}$ is a lattice constant. The eigenergies of the described Hamiltonian, $H_0$, can be found as follows
\begin{eqnarray}
E_{c}(\mathbf k)&=&-\gamma \left |{f(\mathbf k)}\right | ~~,
\nonumber \\
E_{v}(\mathbf k)&=& +\gamma \left |{f(\mathbf k)}\right |~~,
\label{Energy}
\end{eqnarray}
where $c$ and $v$ stand for the CB and VB, respectively.

The coherent electron dynamics in graphene has two major components; one is the intraband dynamics, which is governed by the Bloch acceleration theorem \cite{Bloch_Z_Phys_1929_Functions_Oscillations_in_Crystals},
\begin{equation}
\hbar \frac{{d{\bf{k}}}}{{dt}} = e{\bf{F}}(t),
\label{acceleration}
\end{equation}
with the following solution

\begin{equation}
{{\bf{k}}}({\bf{q}},t) = {\bf{q}} + \frac{e}{\hbar }\int_{ - \infty }^t {{\bf{F}}({t^\prime})d{t^\prime}}. 
\label{kvst}
\end{equation}
where ${\bf q}$ is the initial crystal momentum of an electron in the first BZ.

The solutions of Schr\"odinger equation (\ref{Sch}), for a single band $\alpha$ are the Houston functions \cite{Houston_PR_1940_Electron_Acceleration_in_Lattice},
\begin{equation}
 \Phi^\mathrm{(H)}_{\alpha {\bf q}}({\bf r},t)=\Psi^{(\alpha)}_{\bf{k}(\bf q,t)} ({\bf r})\exp\left(i\phi^{(\mathrm d)}_{\mathrm{\alpha}}({\bf q},t)+i\phi^{(\mathrm B)}_{\mathrm{\alpha}}({\bf q},t)\right),
\end{equation}
where $\alpha=v,c$ stand for the VB and CB, respectively,  $ \mathrm{\Psi^{(\alpha)}_{{\mathbf k}}} $ are Bloch-band eigenstates in the absence of the pulse, and the dynamic phase, $\phi^\mathrm{(D)}_{\mathrm \alpha}$, and geometric phase (Berry phase), $\phi^\mathrm{(B)}_{\mathrm \alpha}$, are defined as  following
\begin{eqnarray}
\phi^\mathrm{(D)}_{\alpha}(\mathbf q,t)= \frac{-1}{\hbar} \int_{-\infty}^t dt^\prime \left(E_\mathrm \alpha[\mathbf k (\mathbf q,t^\prime)]\right),
 \label{phi}
 \\ 
 \phi^\mathrm{(B)}_{\mathrm \alpha}(\mathbf q,t)= \frac{e}{\hbar} \int_{-\infty}^t dt^\prime \mathbf F \left(\mathbfcal{A}^{\mathrm{\alpha \alpha}}[\mathbf k (\mathbf q,t^\prime)]\right),
 \label{phi}
\end{eqnarray}
and  $\mathbfcal{A}^{\alpha\alpha}=\left\langle \Psi^{(\alpha)}_\mathbf q  |   i\frac{\partial}{\partial\mathbf q}|\Psi^{(\alpha)}_\mathbf q   \right\rangle $ is the intraband Berry connection with the following form
\begin{eqnarray}
\mathbfcal{A}^{\alpha\alpha}&=&(\mathcal{A}^{\alpha\alpha}_x,\mathcal{A}^{\alpha\alpha}_y)~,\\
\mathcal{A}_{x}^{\alpha\alpha}(\mathbf k)&=&\frac{-a\gamma ^2}{2\gamma ^2 |f(\mathbf k)|^2}
 \sin \frac{3ak_y}{2\sqrt{3}}\sin{\frac{ak_x}{2}}~,
\nonumber \\
 \label{Ax_alpha}
 \\
\mathcal{A}_{y}^{\alpha\alpha}(\mathbf k)&=&\frac{a\gamma ^2}{2\sqrt{3}\left(\gamma ^2 |f(\mathbf k)|^2\right)} \left(\cos{ ak_x}-\cos{\frac{\sqrt{3}ak_y}{2}}\cos{\frac{ak_x}{2}}\right)~.
 \label{Ay_alpha}
\end{eqnarray}

To find the ultrafast electron dynamics, we solve TDSE (\ref{Sch}) numerically and the solution of this equation can be expanded in the basis of the Houston functions $\Phi^{(H)}_{\alpha {\bf q}}({\bf r},t)$,
\begin{equation}
\Psi_{\bf q} ({\bf r},t)=\sum_{\alpha=c,v}\beta_{\alpha{\bf q}}(t) \Phi^{(H)}_{\alpha {\bf q}}({\bf r},t),
\end{equation}
where 
$\beta_{\alpha{\bf q}}(t)$ are expansion coefficients satisfying the following coupled first order differential equations
\begin{equation}
i\hbar\frac{\partial B_\mathbf q(t)}{\partial t}= H^\prime(\mathbf q,t){B_\mathbf q}(t)~,
\label{Schrodinger}
\end{equation}
where $B_q(t)$ and Hamiltonian, $ H^\prime(\mathbf q,t)$, are defined as 
\begin{eqnarray}
B_\mathbf q(t)&=&\begin{bmatrix}\beta_{c\mathbf q}(t)\\ \beta_{v\mathbf q}(t)\\ \end{bmatrix}~,\\ 
H^\prime(\mathbf q,t)&=&-e\mathbf F(t)\mathbfcal{\hat A}(\mathbf q,t)~,
\end{eqnarray}
where
\begin{eqnarray}
\mathbfcal{\hat A}(\mathbf q,t)&=&\begin{bmatrix}0&\mathbfcal D^{cv}(\mathbf q,t)\\
\mathbfcal D^{vc}(\mathbf q,t)&0\\
\end{bmatrix}~,\\
\mathbfcal D^{cv}(\mathbf q,t)&=&
\mathbfcal A^{cv}[\mathbf k (\mathbf q,t)] \exp\left(i\phi^\mathrm{(D)}_{cv}(\mathbf q,t)+i\phi^\mathrm{(B)}_{cv}(\mathbf q,t)\right),
 \label{Q}
\\
\phi^\mathrm{(D)}_{cv}(\mathbf q,t)&=&\phi^\mathrm{(D)}_{v}(\mathbf q,t)-\phi^\mathrm{(D)}_{c}(\mathbf q,t)
 \label{phiD}
 \\
\phi^\mathrm{(B)}_{cv}(\mathbf q,t)&=&\phi^\mathrm{(B)}_{v}(\mathbf q,t)-\phi^\mathrm{(B)}_{c}(\mathbf q,t)=0
 \label{phiB}
 \\ 
{\mathbfcal{A}}^{cv}({\mathbf q})&=&
\left\langle \Psi^{(c)}_\mathbf q  |   i\frac{\partial}{\partial\mathbf q}|\Psi^{(v)}_\mathbf q   \right\rangle .
\label{D}
\end{eqnarray} 
Here ${\mathbfcal A}^{cv}(\mathbf q)$ is  a matrix element of the non-Abelian Berry connection \cite{Wiczek_Zee_PhysRevLett.52_1984_Nonabelian_Berry_Phase, Xiao_Niu_RevModPhys.82_2010_Berry_Phase_in_Electronic_Properties, Yang_Liu_PhysRevB.90_2014_Non-Abelian_Berry_Curvature_and_Nonlinear_Optics} which can be found analytically as
\begin{eqnarray}
\mathcal{A}_{x}^{cv}(\mathbf k)&=&\Bigg(\frac{-a}{2|f(\mathbf k)|^2}\Bigg)\Bigg( \sin\frac{ak_x}{2}\sin\frac{a\sqrt{3}k_y}{2}\Bigg),
 \label{Ax}
\\
\mathcal{A}_{y}^{cv}(\mathbf k)&=&\Bigg(\frac{a}{2\sqrt{3}|f(\mathbf k)|^2}\Bigg) \Bigg( -1-\cos\frac{a\sqrt{3}k_y}{2}\cos\frac{ak_x}{2}
+2\cos ^2 \frac{ak_x}{2}\Bigg). 
\label{Ay}
\end{eqnarray}

The ultrafast field drives an electric current , ${\mathbf J}(t) = \left\{J_x(t),J_y(t)\right\}$, which can be written as the sum of interband and intraband contributions, $\mathbf J(t)=\mathbf J^\text{(intra)}(t)+\mathbf J^\text{(inter)}(t)$. The intraband current is expressed as 
\begin{equation}
\mathbf J^\text{(intra)}(t)  =\frac{eg_s}{a^2}\sum\limits_{\alpha=\mathrm{c,v},\mathbf q}\left| \beta _{\alpha}(\mathbf q,t) \right|^2\mathbf v^\mathrm{(\alpha )}{(\mathbf k(\mathbf q,t))}~,
\label{intra}
\end{equation}
where $\mathbf v^\mathrm{(\alpha)}({\mathbf  k})=\frac{\partial}{\partial\mathbf k}E^\mathrm{(\alpha)}(\mathbf k)$ is the group velocity or the intraband velocity and $g_s=2$ is the spin degeneracy.
For the known eigenstates, the intraband velocities are calculated as follows

\begin{eqnarray}
\mathrm{ v_x^{ (c)}}(\mathbf k)&=&-\mathrm{v_x^{(v)}}(\mathbf k)=\frac{-a\gamma ^2}{\hbar{|\gamma{f(\mathbf k)}|}}
\sin{\frac{ak_x}{2}}\Big(\cos{\frac{\sqrt{3}ak_y}{2}}
+2\cos{\frac{ak_x}{2}}\Big),\\
\mathrm{v_y^{ (c)}}(\mathbf k)&=&-\mathrm{v_y^{(v)}}(\mathbf k)=\frac{-\sqrt{3}a\gamma ^2}{\hbar{|\gamma{f(\mathbf k)}|}}
\sin{\frac{\sqrt{3}ak_y}{2}}\cos{\frac{ak_x}{2} }.
\end{eqnarray}
The interband current is 
\begin{eqnarray}
   \mathbf J^\text{(inter)}(t)=i\frac{eg_s}{\hbar a^2}\sum _{\substack{\mathbf q\\ \alpha,\alpha^\prime=\mathrm{v,c}\\
 \alpha\ne\alpha^\prime}}\beta _{\alpha^\prime}^\ast(\mathbf q,t)\beta _{\alpha}(\mathbf q,t)\exp \{ i \phi^\mathrm{(D)}_\mathrm{\alpha^\prime\alpha}(\mathbf q,t)\}\left[ E_{\alpha^\prime}\left(\mathbf k(\mathbf q,t)\right)-E_\alpha \left(\mathbf k(\mathbf q,t)\right)\right] \mathbfcal A^{(\alpha\alpha^\prime)}\left(\mathbf k(\mathbf q,t)\right),
 \label{J}
\end{eqnarray}
where 
\begin{eqnarray}
&&\phi^\mathrm{(D)}_\mathrm {\alpha^\prime \alpha}(\mathbf q,t)=\phi^\mathrm{(D)}_ \mathrm {\alpha}(\mathbf q,t)-\phi^\mathrm{(D)}_ \mathrm {\alpha^\prime }(\mathbf q,t).
\label{Eq:interband Berry phase}
\end{eqnarray}

\section{Results and Discussion}

\begin{figure}
\begin{center}\includegraphics[height=9cm,width=10cm]{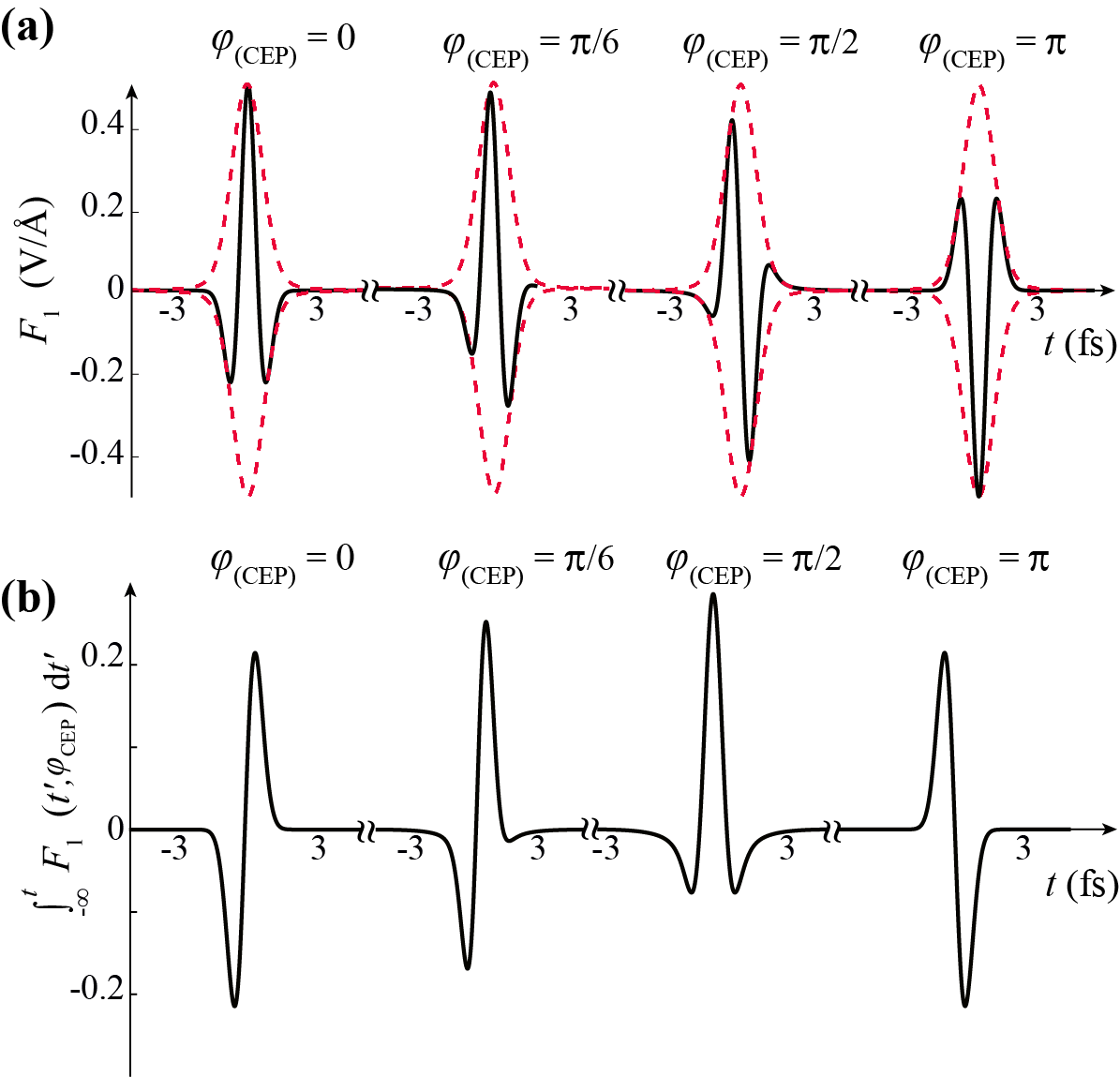}\end{center}
  \caption{(Color online) (a) For different $\varphi_\mathrm{CEP}$ values the dash line shows the envelop and the solid black line shows waveform of $F_1(t)$. (b) The integral of the field $F_1(t)$ which governs the electron intraband dynamics is illustrated for different $\varphi_\mathrm{CEP}$ values. The amplitude of the field is $0.5~\mathrm{V/\AA}$}
  \label{fig:H2_p_int_CEP}
\end{figure}

\begin{figure}
\begin{center}\includegraphics[height=9cm,width=10cm]{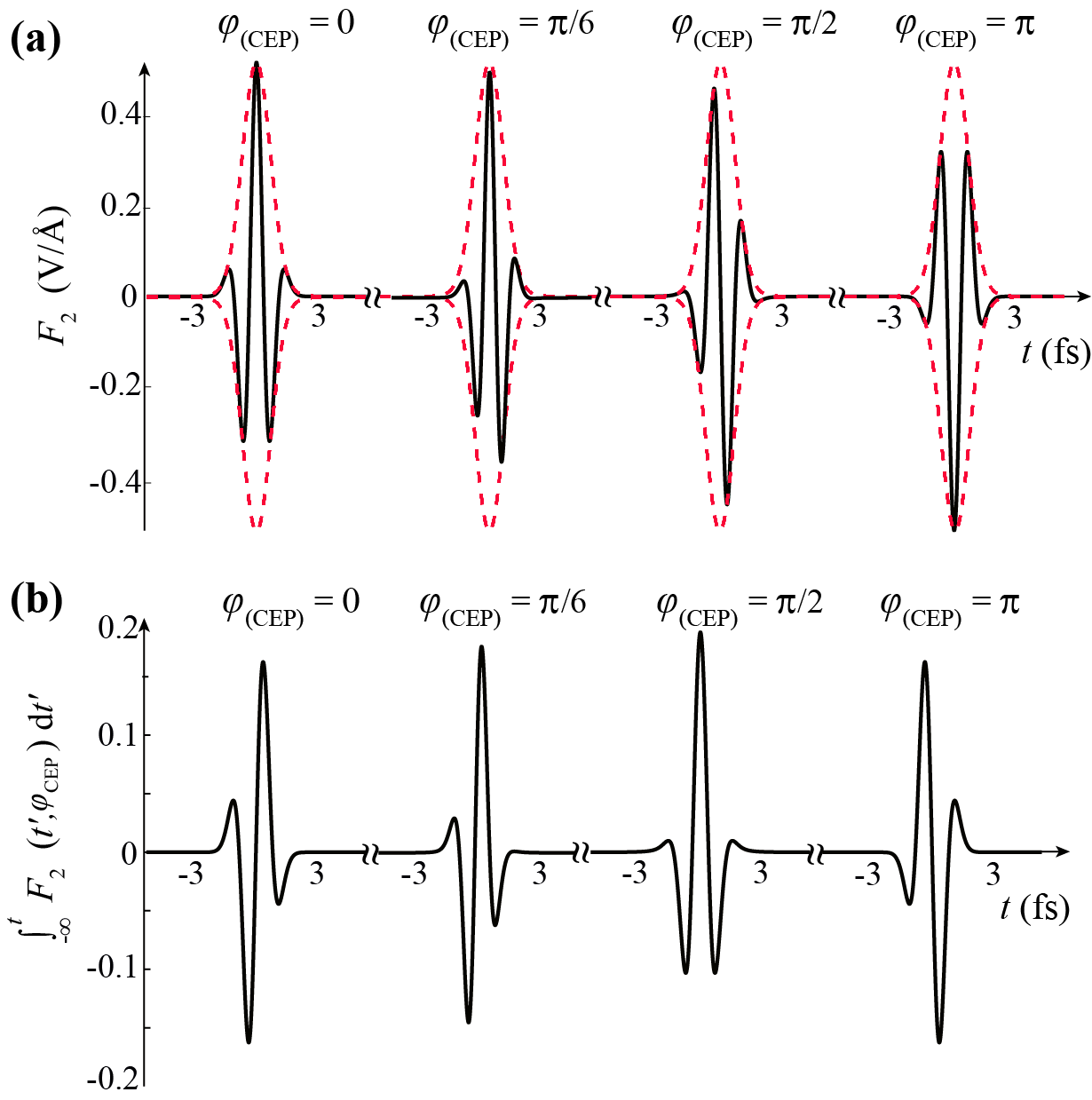}\end{center}
  \caption{(Color online) The same as in Fig. \ref{fig:H2_p_int_CEP}, but for field $F_2(t)$. }
  \label{fig:H4_p_int_CEP}
\end{figure}

We consider a linearly polarized pulse with the following waveform, $F_{n}(t,\varphi_\mathrm{CEP})$,
\begin{eqnarray}
&&F_n(t,\varphi_\mathrm{CEP})=A_n(t)\cos \left(\phi(t)+\varphi_\mathrm{CEP}\right)\\
&&A_n(t)=\sqrt{f^2_n(t)+\tilde{f}^2_n(t)}\\
&&\phi(t)=\arctan \left(\frac{\tilde{f}(t)}{f(t)}\right)
\end{eqnarray}
where $\varphi_\mathrm{CEP}$ is the carrier-envelope phase, n is 1 or 2, $\tilde{f}_n(t)$ is Hilbert transformation of $f_n(t)$, and $f_1(t)$ and $f_2(t)$ are defined as following Hermite-Gaussian expressions
\begin{eqnarray}
&&f_1(t)=F_0\mathrm{H^{(2)}}(u)\exp{\left(-u^2\right)}\\
&&f_2(t)=F_0\mathrm{H^{(4)}}(u)\exp(-u^2),
\end{eqnarray} 
 where $F_0$ is the amplitude of the field, $\mathrm{H^{(2)}}(u)$ and $\mathrm{H^{(4)}}(u)$ are Hermite polynomials \cite{Thompson_Book_1886_Hermite_polynomial} of degree 2 and 4, respectively,
\begin{eqnarray}
&&\mathrm{H^{(2)}}(u)=-2u^2+1\nonumber\\
&&\mathrm{H^{(4)}}(u)=\frac{1}{12}(16u^4-48u^2+12),\nonumber
\end{eqnarray}
  $u=\frac{t}{\tau}$, and $\tau=1~\mathrm{fs}$.
The carrier-envelope phase, $\varphi_\mathrm{CEP}$, is defined as the phase difference between the carrier field  and the envelope, which are shown with solid black lines and dash red lines, respectively, in Figs.\ \ref{fig:H2_p_int_CEP} (a) and \ref{fig:H4_p_int_CEP} (a).
 The phase,$\varphi_\mathrm{CEP}$,  modifies the waveform of the field without changing the bandwidth of the pulse. Figures \ref{fig:H2_p_int_CEP}(a) and \ref{fig:H4_p_int_CEP}(a) show different waveforms of two fields, $F_1(t)$ and $F_2(t)$, for different $\varphi_\mathrm{CEP}$. Here $F_2(t)$ has more oscillation in comparison to $F_1(t)$. As shown in panels (b) of these figures, the whole integral of the pulse, known as the vector potential, is zero for all  $\varphi_\mathrm{CEP}$s, which means that the pulse has zero area. Applying zero area pulse causes that the electron gets back to its initial crystal momentum at the end of the pulse according to Eq.\ (\ref{kvst}).

\begin{figure}
\begin{center}\includegraphics[height=4cm,width=6cm]{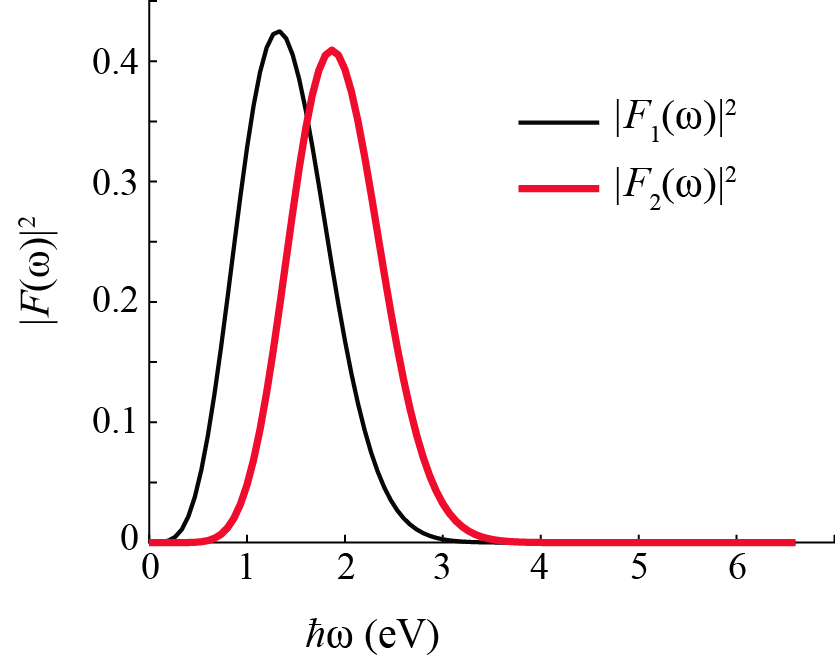}\end{center}
  \caption{(Color online) Fourier transform of $F_1(t,\varphi_\mathrm{CEP})$ and $F_2(t,\varphi_\mathrm{CEP})$ are shown with black and red lines respectively. Since $F_2(t,\varphi_\mathrm{CEP})$ has more oscillations its bandwidth is shifted to higher energies. The mean frequencies, $\overline{\omega}$, of $1.40~\mathrm{eV}/\hbar$ and $1.92~\mathrm{eV/}\hbar$ are for $|F_1(\omega)|^2$ and $|F_2(\omega)|^ 2$ respectively. }
  \label{fig:FFT_H2_H4}
\end{figure}

The applied pulses $F_1(t)$ and $F_2(t)$ have different number of oscillations and correspondingly different bandwidth, where Fig.\ \ref{fig:FFT_H2_H4} shows Fourier transforms of two pulses in the energy domain with the mean frequencies, $\overline{\omega}$, of $1.40~\mathrm{eV}/\hbar$ and $1.92~\mathrm{eV/}\hbar$ respectively. The mean frequency is calculated as
\begin{equation}
\overline{\omega}=\frac{\int \omega \left|\mathbf{F}_{n}(\omega)\right|^{2} d \omega}{\int \left|\mathbf{F}_{n}(\omega)\right|^{2} d \omega}.
\end{equation}

\begin{figure}
\begin{center}\includegraphics[height=6cm,width=12cm]{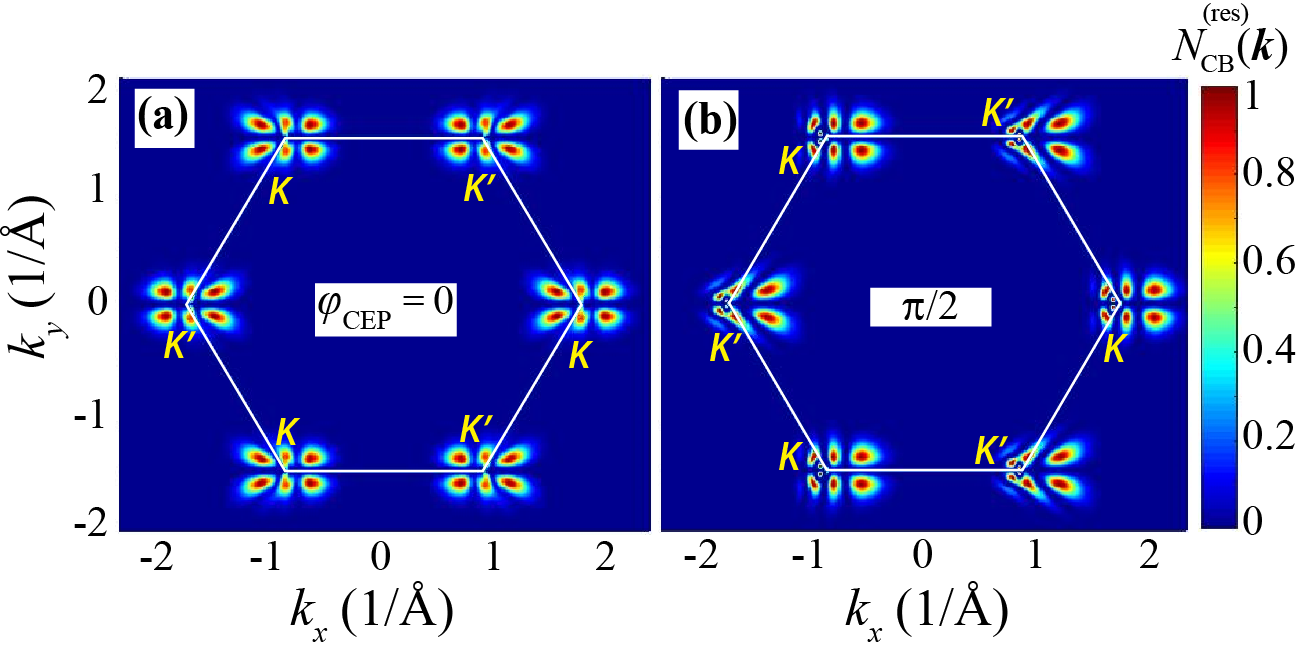}\end{center}
  \caption{(Color online) The distributions of the residual CB population, $N\mathrm{^{(res)}_{CB}}(\mathbf k)$, are illustrated for (a) $\varphi_\mathrm{CEP}=0$ and (b) $\varphi_\mathrm{CEP}=\pi/2$. The applied field is $F_1(t)$ with the amplitude of $0.5~\mathrm{V/\AA}$. The white solid lines show the edges of the first BZ and Dirac points, $K$ and $K^\prime$, are indicated on its corners.}
  \label{fig:CB_F0_0p5_H2_H3}
\end{figure}%

We assume that before applying the pulse the conduction band is empty and valence band is full. The applied ultrafast pulse with high amplitude excites the electrons from VB to CB and redistributes the electrons in CB and VB within a few femtoseconds. After the pulse ends, there remain some electron population in CB which is called the residual CB population. 
The nonzero residual population of CB shows the irreversibility of the ultrafast electron dynamics in graphene. The irreversibility of the electron dynamics in graphene is due to zero energy gap and correspondingly strong interband coupling at its Dirac points.   The distribution of the residual CB population for two different values of $\varphi_\mathrm{CEP}$, 0 and $\frac{\pi}{2}$ for  applied field $F_1(t)$ are shown in Fig.\ \ref{fig:CB_F0_0p5_H2_H3} panels (a) and (b) respectively.
As shown in panel (a) the electron distribution in the extended BZ is symmetric with respect to both x and y-axes. The reason for the presence of the symmetry with respect to the y-axis is that the integral of the field or the vector potential has a symmetric profile for $\varphi_\mathrm{CEP}$ equals to 0 as shown in Fig.\ \ref{fig:H2_p_int_CEP}(b). However due to the asymmetric profile of the vector potential for $\varphi_\mathrm{CEP}$ equals to $\pi/2$ shown in Fig.\ \ref{fig:H2_p_int_CEP} (b), the distribution of electron, see Fig.\ \ref{fig:CB_F0_0p5_H2_H3} (b), is not symmetric with respect to y-axis, e.g., right sides of the Dirac points are more populated than the left sides.

\begin{figure}
\begin{center}\includegraphics[height=6cm,width=12cm]{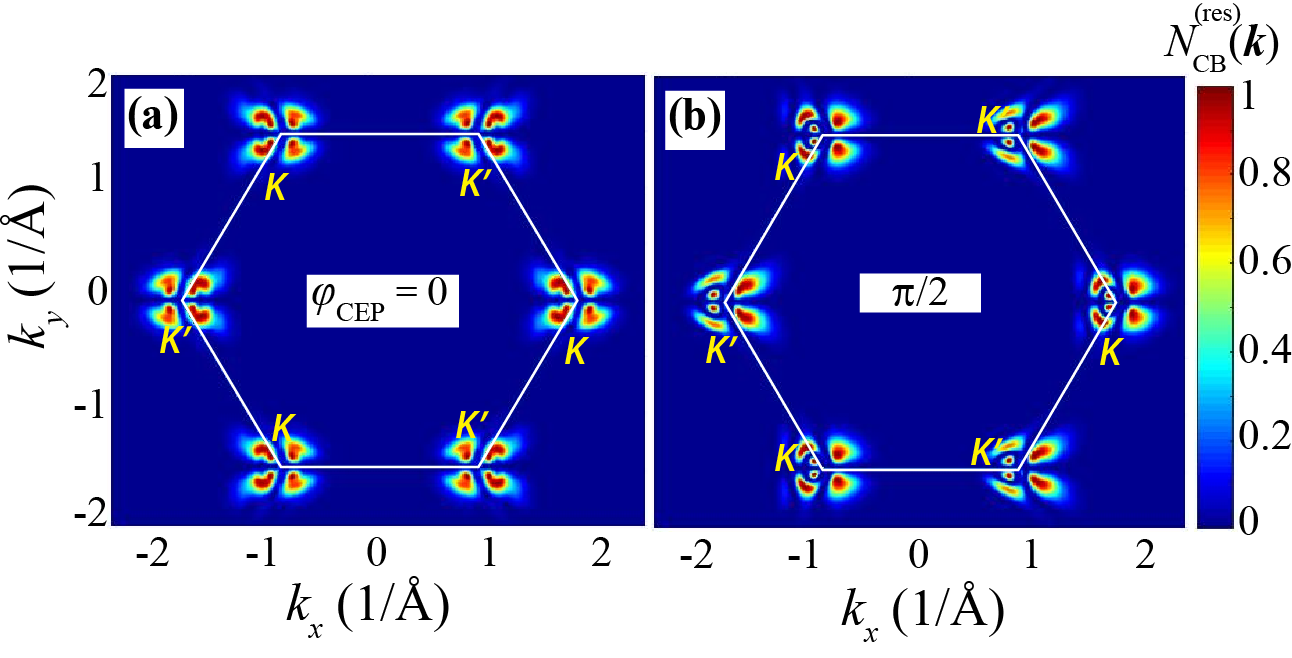}\end{center}
  \caption{(Color online) . The same as in Fig.\ \ref{fig:CB_F0_0p5_H2_H3}, but for applied field $F_2(t)$.}
  \label{fig:CB_F0_0p5_H4_H5}
\end{figure}%

As illustrated in Fig.\ \ref{fig:CB_F0_0p5_H4_H5}, by applying $F_2(t)$ the number of hot spots is decreasing compared to the distribution of residual CB population induced by $F_1(t)$, shown in in Fig.\ \ref{fig:CB_F0_0p5_H2_H3} . 
As it is manifested in Fig.\  \ref{fig:CB_F0_0p5_H4_H5}, similar to Fig.\ \ref{fig:CB_F0_0p5_H2_H3} and for  $\varphi_\mathrm{CEP}$ equals to zero, there is no differences in the population of two valleys $K$ and $K^\prime$ however for  $\varphi_\mathrm{CEP}$ equals to $\pi/2$ the difference in the distribution of the electron in two valleys leads to the valley polarization, which will be studied somewhere else.
The electron population in the first BZ of graphene can be observed by using time-resolved angle-resolved photoemission spectroscopy (tr-ARPES)\cite{Freericks_et_al_Annal_Phys_2017_Superconductors_TR-ARPES_Theory,Chiang_et_al_PhysRevLett.107_Berry_Phase_in_Graphene_ARPES} with high temporal resolution.
\begin{figure}
\begin{center}\includegraphics[height=5cm,width=6cm]{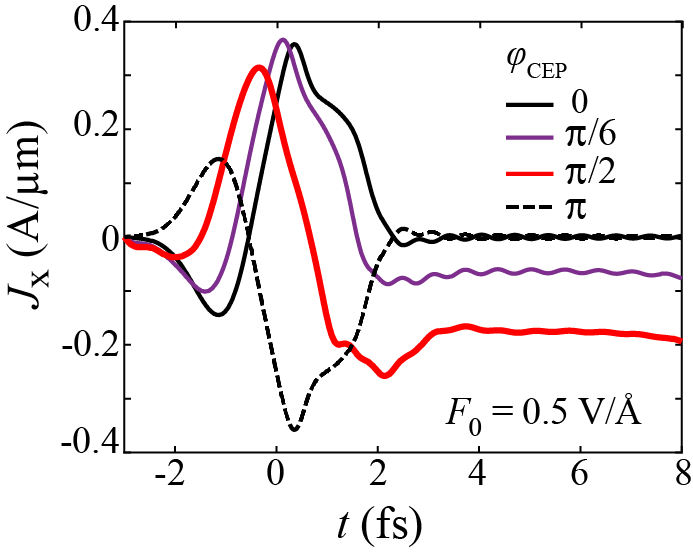}\end{center}
  \caption{(Color online) The charge current density as a function of time are calculated for different values of $\varphi_\mathrm{CEP}$, 0, $\pi/6$, $\pi/2$, and $\pi$. The applied field is $F_1(t,\varphi_\mathrm{CEP})$ with the amplitude of $0.5~\mathrm{V/\AA}$.  }
  \label{fig:H2_H3_current_Fx_0p5_gap_0}
\end{figure}%

The applied ultrashort field redistribute the electron in VB and CB, and drives an ultrafast current. This current is a result of intraband and interband motions of an electron in graphene, see Eqs. (\ref{intra}) and (\ref{J}). As illustrated in Fig.\ \ref{fig:H2_H3_current_Fx_0p5_gap_0}, for $\varphi_\mathrm{CEP}=0$ , the profile of the current follows the vector potential of the field, shown in Fig.\ \ref{fig:H2_p_int_CEP}(b), and the residual current after the pulse is zero. This is due to the symmetric profile of the vector potential, which makes a symmetric distribution of the residual CB population, illustrated in Fig.\ \ref{fig:CB_F0_0p5_H2_H3}. However, there is nonzero residual currents for $\varphi_\mathrm{CEP}=\pi/6$ and $\varphi_\mathrm{CEP}=\pi/2$, which are due to asymmetric profiles of the vector potentials, see Fig.\ \ref{fig:H2_p_int_CEP} (b), and as a result the asymmetric distributions of the residual CB population, see Fig.\ \ref{fig:CB_F0_0p5_H2_H3} (b). For  $\varphi_\mathrm{CEP}=\pi$, the current has the same amplitude as the current for $\varphi_\mathrm{CEP}=0$ but it propagates in opposite direction since the field changes its direction to the opposite one for $\varphi_\mathrm{CEP}=\pi$.

\begin{figure}
\begin{center}\includegraphics[height=5cm,width=13cm]{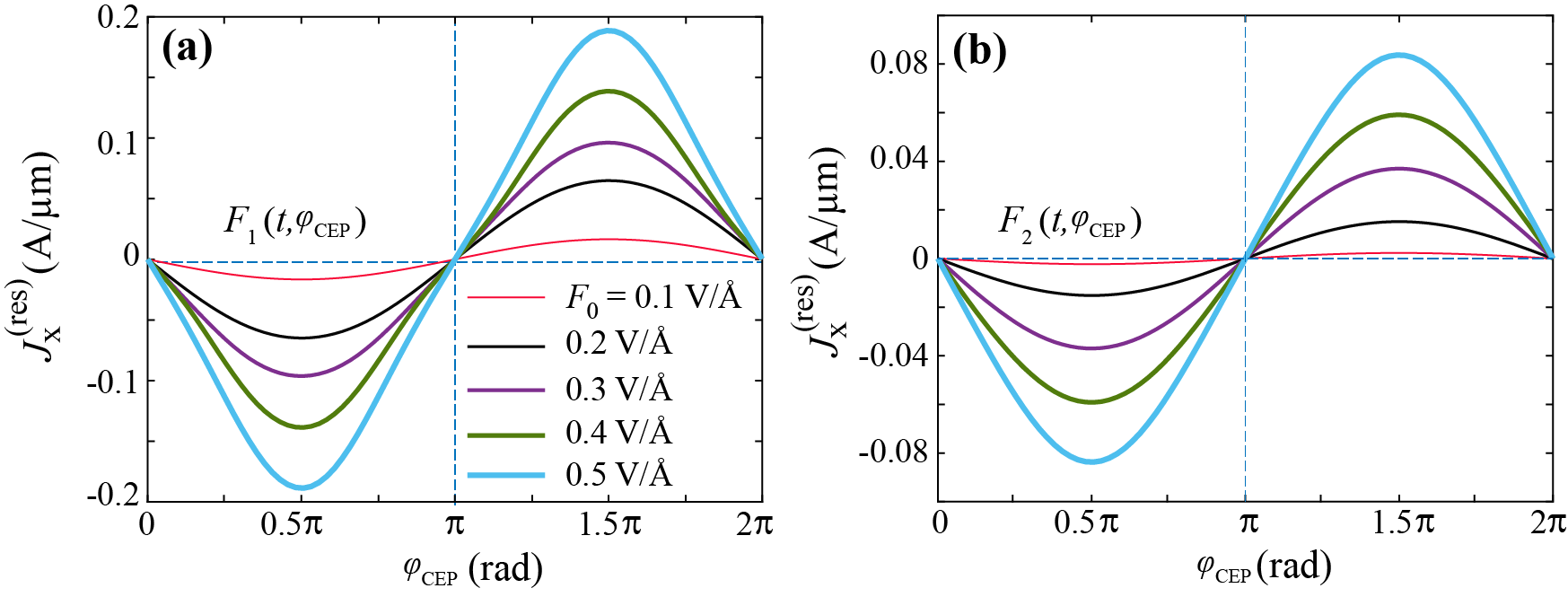}\end{center}
  \caption{(Color online) The residual currents are shown as functions of $\varphi_\mathrm{CEP}$.The applied field has a linear polarization in the x direction with the different field amplitudes, $0.1~\mathrm{V/\AA}$ - $0.5~\mathrm{V/\AA}$. (a) and (b) show the residual currents for the applied pulses which carries different mean frequencies. The legends are shown in panel (a).}
  \label{fig:H2_H4_residual_intra_Jx_gap_0}
\end{figure}%

Figure \ref{fig:H2_H4_residual_intra_Jx_gap_0} illustrates the residual current densities as a function of $\varphi_\mathrm{CEP}$ for different values of field amplitudes, $F_0$,  and for the applied fields $F_1(t)$ and $F_2(t)$.
The residual current densities oscillate similar to a sinusoidal function. Their amplitudes increase with the field amplitudes. Also, the residual current density for field $F_2(t)$ is less than the residual current density for field $F_1(t)$ because there are more oscillations in field $F_2(t)$.
\begin{figure}
\begin{center}\includegraphics[height=5cm,width=13cm]{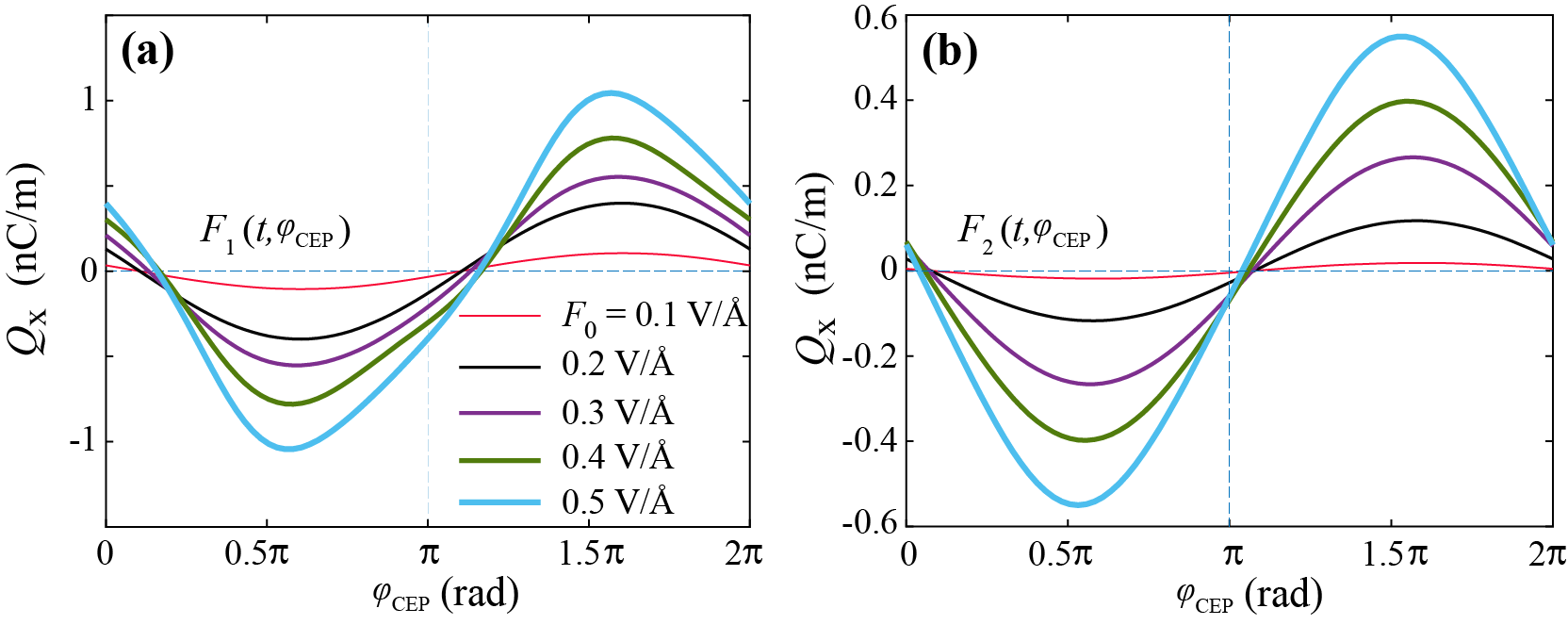}\end{center}
  \caption{(Color online) The transferred charges per unit length are shown as functions of the CEP.The applied field has a linear polarization in the x direction with the different field amplitudes, $(0.1-0.5~\mathrm{V/\AA})$. (a) and (b) show the densities of transferred charge for the applied pulses which carries different mean frequencies. The legends are shown in panel (a).}
  \label{fig:H2_H4_trans_Qx_gap_0}
\end{figure}%

Due to the nonlinear process even for the case of $\varphi_\mathrm{CEP}=0$ where the residual current is zero, see Fig.\ \ref{fig:H2_H3_current_Fx_0p5_gap_0}, the profile of the ultrafast field driven current has a nonzero area which results in a nonzero transferred charge density. The transferred charge can be measured experimentally, and it characterizes the residual electric polarization of the graphene induced by the ultrafast field. For the case of $\varphi_\mathrm{CEP}=0$, the electric charge is transferred during the pulse propagation; however, for the cases of $\varphi_\mathrm{CEP}=\pi/6$ and $\varphi_\mathrm{CEP}=\pi/2$ some electric charge transfers not only during the pulse but also after the end of the pulse due to the nonzero residual current. Figure.\ \ref{fig:H2_H4_trans_Qx_gap_0} shows the transferred charge density, $Q_\mathrm{x}$ for both applied fields, $F_1(t)$ and $F_2(t)$ respectively in planes (a) and (b).
By comparing two panels (a) and (b) we can see that the amount of the transferred charge is decreasing with increasing the number of oscillations in the pulse.

\section{Conclusion}
We studied the effect of the ultrafast laser pulse waveform on the dynamics of massless Dirac fermions in graphene numerically. We modified the waveform of the applied pulse by the carrier-envelope phase and also by two different Hermit Gaussian polynomials which are used to describe the profile of the field. We calculated the ultrafast field-driven current for different carrier-envelope phases. Depending on the carrier-envelope phase, the current propagates even after the pulse ends, which is called the residual current.  The unbalanced profile of the current density results in the nonzero transferred charge density, which defines the final polarization of graphene. The residual current and the transferred charge densities are decreasing with increasing the number of the oscillation in the pulse for a fixed carrier-envelope phase and a fixed field amplitude. Our results illustrate that the ultrafast nonlinear electron dynamics is affected by the waveform of the applied ultrashort pulse. The distribution of the electron population in the reciprocal space can be observed experimentally by time- and angle-resolved photoemission spectroscopy (tr-ARPES)\cite{Freericks_et_al_Annal_Phys_2017_Superconductors_TR-ARPES_Theory,Chiang_et_al_PhysRevLett.107_Berry_Phase_in_Graphene_ARPES}.

\begin{acknowledgments}
Major funding was provided by Grant No. DE-FG02-11ER46789 from the Materials Sciences and Engineering Division of the Office of the Basic Energy Sciences, Office of Science, U.S. Department of Energy. Numerical simulations have been performed using support by
Grant No. DE-FG02-01ER15213 from the Chemical Sciences, Biosciences and Geosciences Division, Office of Basic Energy Sciences, Office of Science, US Department of Energy. The work of V.A. was supported by NSF EFRI NewLAW Grant EFMA-17 41691. Support for S.A.O.M. came from a MURI Grant No. FA9550-15-1-0037 from the US Air Force of Scientific Research.
\end{acknowledgments}


\end{document}